\documentclass{article-hermes}	

\usepackage[utf8x]{inputenc}
\usepackage[T1]{fontenc}
\usepackage{pstricks}

\title{Segmentation en locuteurs de séries \textsc{tv}}

\author{Xavier Bost\fup{*} \andauthor Georges Linarès\fup{*}  } 

\address{%
\fup{*} Laboratoire Informatique d'Avignon
}

\resume{La segmentation de flux audio en locuteurs apparaît
  particulièrement délicate lorsqu'elle est appliquée à des films de
  fiction, où de nombreux personnages parlent dans des conditions
  acoustiques variables (musique de fond, bruitages, fluctuations dans
  l'intonation...). Au-delà d'une telle variabilité acoustique, ce
  type de films exhibe cependant de la régularité sur le plan visuel,
  particulièrement dans les passages dialogués. Nous introduisons dans
  ce papier une méthode en deux temps pour procéder à la segmentation
  en locuteurs d'épisodes de séries \textsc{tv}~: un premier
  regroupement en locuteurs est effectué localement, dans les limites
  de scènes visuellement identifiées comme des dialogues~; les
  locuteurs conjecturés sont ensuite comparés lors d'une deuxième
  phase de regroupement afin de détecter les locuteurs récurrents~:
  cette deuxième étape de regroupement a lieu sous la contrainte que
  les différents locuteurs impliqués dans un même dialogue soient
  assignés à des groupes distincts. Les performances obtenues en
  suivant cette approche sont comparées à celles qu'on obtient en
  appliquant aux mêmes données des outils standards de segmentation en
  locuteurs.}

\abstract{Speaker diarization of audio streams turns out to be
  particularly challenging when applied to fictional films, where many
  characters talk in various acoustic conditions (background music,
  sound effects, variations in intonation...). Despite this acoustic
  variability, such movies exhibit specific visual patterns,
  particularly within dialogue scenes. In this paper, we introduce a
  two-step method to achieve speaker diarization in \textsc{tv}
  series: speaker diarization is first performed locally within scenes
  visually identified as dialogues; then, the hypothesized local
  speakers are compared to each other during a second clustering
  process in order to detect recurring speakers: this second stage of
  clustering is subject to the constraint that the different speakers
  involved in the same dialogue have to be assigned to different
  clusters. The performances of our approach are compared to those
  obtained by standard speaker diarization tools applied to the same
  data.}

\motscles{
  Segmentation en locuteurs$_1$,
  Structuration de vidéos$_2$,
  regroupement non supervisé$_3$.
}


\begin{document}

\maketitlepage

\section{Introduction}

\label{sec:intro}

La recherche d'information dans les flux audiovisuels rencontre une
difficulté majeure liée à l'absence fréquente de métadonnées décrivant
les contenus et la façon dont ils sont structurés \cite{Quenot10},
\cite{Lew06}. Extraire l'organisation sous jacente des flux de
contenus est donc une étape critique du processus d'indexation, qui
comporte deux volets. D'une part, il faut segmenter le flux continu en
un ensemble de segments homogènes. D'autre part, il faut caractériser
les segments extraits par des descripteurs qui peuvent être de natures
très diverses \cite{adams03}.  Il peut s'agir de descripteurs visuels
pour la segmentation en plans ou en scènes, de descripteurs des
contenus sémantiques pour une segmentation thématique, des
descripteurs caractérisant le style éditorial d'un segment dans la
segmentation en genres vidéo... De nombreuses applications considèrent
les contenus parlés et l'identité des locuteurs comme un élement
essentiel à la structuration. Dans cet article, nous nous intéressons
à la segmentation en locuteurs dans le cadre, assez général, de
l'indexation de grandes bases audiovisuelles.

La détection des locuteurs impliqués dans des contenus audio est
usuellement conçue comme une tâche de segmentation du flux, qui
consiste à assigner les différents segments parlés à leurs locuteurs
respectifs. Cette tâche est donc habituellement accomplie en deux temps,
simultanés ou successifs~: détection des points de changement entre
locuteurs~; regroupement des segments résultants en classes de
locuteurs. Ce regroupement est souvent mené en appliquant un
algorithme de regroupement hiérarchique, soit ascendant, soit
descendant~\cite{evans2012comparative}.

Le regroupement des segments en classes de locuteurs est mené de
manière non supervisée~: le nombre total de locuteurs en particulier
n'est pas \textit{a priori} fixé et le partitionnement optimal des
segments parlés en classes de locuteurs doit être automatiquement
déterminé.

Les systèmes de segmentation en locuteurs ont d'abord été conçus pour
traiter des flux audio spécifiques dans des conditions acoustiques
défavorables, mais bien balisées~: conversations téléphoniques,
journaux d'information, réunions de travail. Des travaux récents ont
appliqué ces systèmes aux contenus vidéos, dont le contexte de
production, moins bien défini, induit une plus forte variabilité.

Dans \cite{clement2011speaker}, les auteurs appliquent des outils
standards de segmentation en locuteurs à la source audio de documents
vidéos de différents genres collectés sur le web. Les performances
mesurées se dégradent sensiblement pour les dessins animés et les
bandes-annonces de films, avec des taux d'erreur (Diarization Error
Rate) élevés~: parmi les raison
incriminées, les auteurs mentionnent le nombre élevé de locuteurs
impliqués dans ces flux, ainsi que de fortes variabilités dans
l'environnement acoustique (superposition de la parole et de la
musique de fond, bruitages...).

De même, les taux d'erreurs reportés
dans \cite{ercolessi2013extraction} après application d'un outil
standard de segmentation en locuteurs au canal audio de quatre
épisodes d'une série \textsc{tv} apparaissent très élevés
(\textsc{der} $\simeq 70\%$).

Récemment, plutôt que d'aborder les flux vidéo par le seul canal
audio, certains travaux se sont concentrés sur des approches
multimodales pour procéder à la segmentation en locuteurs~: dans
\cite{Friedland2009}, les auteurs évaluent une méthode fondée sur une
fusion précoce de mélanges de gaussiennes audio et vidéo suivie de
l'application d'un algorithme agglomératif au flux bicanal qui en
résulte. Cette technique est évaluée sur le corpus \textsc{ami}
\cite{Carletta2005}, constitué d'enregistrements audiovisuels de
quatre participants jouant des rôles dans un scénario de réunion.

Appliquer un système de segmentation en locuteurs à des épisodes de
séries télévisées, où le nombre de locuteurs est globalement plus
élevé que dans des longs métrages, peut donc sembler particulièrement
délicat. Toutefois, les films de fiction exhibent sur le plan visuel
de nombreuses régularités formelles. Le montage des scènes dialoguées
en particulier exige que la règle des \og{}180 degrés\fg{} soit
respectée afin que deux interlocuteurs filmés alternativement donnent
l'impression de se regarder en se parlant~: pour suggérer le
croisement des regards, l'un doit regarder vers la droite de l'écran
et l'autre vers la gauche. Les deux caméras qui les filment doivent
donc se situer du même côté d'une ligne imaginaire qui les relierait
l'un à l'autre. L'observation d'une telle règle induit un motif
d'alternance entre deux plans récurrents, caractéristique des scènes
dialoguées.

Sur la base de ces régularités formelles, nous proposons ici de
décomposer le processus de segmentation en locuteurs en deux étapes
lorsqu'il est appliqué aux films de fiction~: une première passe de
regroupement des segments en classes de locuteurs est effectuée
localement, dans les limites de courtes scènes visuellement détectées
comme des dialogues~; une seconde étape vise à détecter les locuteurs
récurrents d'une scène à l'autre en procédant à un second regroupement
des classes de locuteurs conjecturées localement~; cette seconde phase
de regroupement est opérée sous la contrainte que les locuteurs
impliqués dans une même scène ne puissent pas être regroupés dans la
même classe.

Un tel regroupement en deux temps est étroitement apparenté à ce qui
est désigné dans \cite{tran2011comparing} sous le terme
d'\og{}architecture hybride\fg{} dans un contexte de segmentation en
locuteurs d'émissions successives~: les locuteurs des différentes
émissions sont d'abord détectés indépendamment, avant que les
locuteurs récurrents d'une émission à l'autre ne soient
regroupés. Dans \cite{bendris2013unsupervised}, les auteurs
s'efforcent de combiner segmentation en locuteurs et regroupement des
visages pour procéder à l'identification des personnes impliquées dans
un débat filmé~: la meilleure des deux modalités pour identifier une
personne est retenue, avant que l'information acquise ne soit propagée
aux autres éléments de la classe associée. Enfin, la segmentation en
locuteurs a été appliquée aux séries \textsc{tv}, mais plutôt comme une
méthode, parmi d'autres sources, de segmentation du flux en scènes
présentant une unité narrative. Dans \cite{bredin2012segmentation},
les performances obtenues par des approches mono-modales et
multi-modales pour procéder à la segmentation en scènes sont évaluées
et comparées.

Dans ce papier, plutôt que d'utiliser la segmentation en locuteurs
pour structurer le film, on propose de s'appuyer sur sa structure
narrative, telle qu'on peut l'inférer à partir d'indices visuels,
pour améliorer la segmentation en locuteurs de tels contenus. La façon
dont les scènes dialoguées sont visuellement détectées est décrite
dans la section~\ref{sec:dialogues}~; les deux étapes de regroupement
en locuteurs sont introduites dans les sections~\ref{sec:seg_locale}
et~\ref{sec:seg_globale}~; les résultats expérimentaux obtenus sont
donnés dans la section~\ref{sec:expe}.

\section{Détection visuelle de scènes dialoguées}

\label{sec:dialogues}

Le flux vidéo peut être globalement considéré comme une suite finie
d'images fixes projetées sur l'écran à un rythme constant propre à
simuler pour l'\oe il humain la continuité du mouvement. Par ailleurs,
selon \cite{koprinska2001temporal}, un plan constitue une unité
définie comme une \og{}suite ininterrompue d'images prises par une
caméra unique\fg{}.

Comme on l'a relevé dans la section~\ref{sec:intro}, les scènes
dialoguées sont caractérisées par des motifs spécifiques où des plans
récurrents alternent.

Détecter ces motifs caractéristiques des dialogues exige donc d'une
part de segmenter le flux vidéo en plans et d'autre part de
déterminer les plans semblables.

\subsection{Segmentation en plans et détection des plans semblables}

Défini par la continuité des images qu'il contient, un plan peut aussi
être défini en opposition aux images des plans qui lui sont
temporellement contigus. La segmentation en plans est donc usuellement
conçue comme une tâche de détection des ruptures, soit brutales
(coupes), soit graduelles (fondus), dans la continuité des images
constitutives du flux vidéo. Marginales dans les trois séries
\textsc{tv} incluses dans notre corpus, les transitions graduelles
sont ici ignorées.

Une coupe entre plans est donc détectée dès que deux images
consécutives diffèrent sensiblement. De même, deux plans sont
considérés comme semblables si la dernière image du premier et la
première image du second sont suffisamment proches.

Détecter les coupes entre plans et repérer les plans similaires
suppose donc que la similarité entre deux images puisse être
évaluée. A cette fin, chaque image est décrite par la distribution
statistique des pixels qu'elle contient dans l'espace colorimétrique
\textsc{hsv} (teinte, saturation, luminance) et la distance entre deux
images est mesurée par la corrélation de leurs histogrammes
colorimétriques respectifs. Toutefois, il n'est pas exclu que deux
images différentes partagent le même histogramme colorimétrique~: afin
de prévenir de tels rapprochements erronés, l'information de
localisation des couleurs sur l'image est réintroduite en divisant
chaque image en 30 blocs, chacun étant associé à son propre
histogramme de couleurs~; les images sont alors comparées bloc par
bloc en mesurant la corrélation des histogrammes associés selon la
méthode décrite dans~\cite{koprinska2001temporal}.

Les deux seuils de corrélation utilisés pour détecter les coupes et
les plans semblables ont été estimés à partir d'expériences sur un
sous-ensemble de développement.

\subsection{Détection des passages dialogués}

\label{ssec:motif}

Soit $\Sigma = \{l_1, ..., l_m\}$ un ensemble de $m$ étiquettes de
plans, deux plans partageant la même étiquette s'ils sont conjecturés
comme semblables au sens de la sous-section précédente.

En assimilant un plan à son étiquette, un film peut alors être décrit
comme une chaîne finie de plans $\mathbf{s} = s_1s_2...s_k$ ($s_i
\in \Sigma$).

Pour tout couple de plans $(l_1, l_2) \in \Sigma^2$, l'expression
régulière $r(l_1, l_2)$ qui suit désigne une partie de l'ensemble de
toutes les suites possibles de plans $\Sigma^* = \bigcup_{n \geqslant
  0} \Sigma^n$~:

\begin{equation}
  r(l_1, l_2) = \Sigma^* l_1(l_2l_1)^+ \Sigma^*
  \label{eq:patt1}
\end{equation}

L'ensemble $\mathcal{L}(r(l_1, l_2))$ des chaînes désignées par
l'expression régulière~\ref{eq:patt1} correspond alors à toutes les
suites de plans où le plan $l_2$ est inséré entre deux occurrences du
plan $l_1$, avec une éventuelle répétition de la séquence $(l_2l_1)$,
quels que soient les plans qui précèdent et suivent cette
séquence. Cette expression régulière vise à capturer les suites de
plans alternants caractéristiques des passages dialogués.

\begin{figure}[htb]
  \vspace{0.5cm}
  \begin{minipage}[b]{1.0\linewidth}
    \centering
    \centerline{\includegraphics[width=9cm]{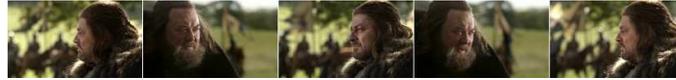}}
  \end{minipage}
  \caption{{\it Exemple de séquence de plans $...l_1l_2l_1l_2l_1...$
      capturée par l'expression régulière~\ref{eq:patt1} pour les deux
      étiquettes de plans $l_1$ et $l_2$.}}
  \vspace{0.5cm}
  \label{fig:motif_1}
\end{figure}

La Figure~\ref{fig:motif_1} présente le type de séquence de plans
capturée par l'expression régulière $r(l_1, l_2)$ et illustre la
règle des \og{}180 degrés\fg{} évoquée dans la
section~\ref{sec:intro}~: les deux personnages regardent dans des
directions opposées.

Pour un film caractérisé par la séquence de plans $\mathbf{s} =
s_1s_2...s_k$ ($s_i \in \Sigma$), on définit l'ensemble
$\mathcal{P}(\mathbf{s}) \subseteq \Sigma^2$ de tous les couples de
plans impliqués dans les séquences de plans alternants dénotées par
l'expression~\ref{eq:patt1}~:

\begin{equation}
  \mathcal{P}(\mathbf{s}) = \{(l_1, l_2) \ | \ \mathbf{s} \in
  \mathcal{L}(r(l_1, l_2))\}
\end{equation}

Pour $(l_1, l_2) \in \mathcal{P}(\mathbf{s})$, l'ensemble
$\mathbf{u}(l_1, l_2)$ regroupe alors tous les segments parlés
couverts par les séquences où les plans $l_1$ et $l_2$ alternent selon
les modalités définies par la règle~\ref{eq:patt1}.

\begin{figure}[htb]
  \vspace{0.5cm}
  \begin{minipage}[b]{1.0\linewidth}
    \centering
    \centerline{\includegraphics[width=9cm]{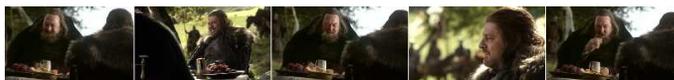}}
  \end{minipage}
  \caption{{\it Séquence de plans $...l_1l_2l_1l_3l_1...$ à la
      frontière de deux motifs adjacents $(l_1, l_2)$ et $(l_1, l_3)$
      avec un plan en commun.}}
  \vspace{0.5cm}
  \label{fig:motif_2}
\end{figure}

Afin d'accroître la couverture des motifs de $\mathcal{P}(\mathbf{s})$
tout en réduisant leur dispersion, deux extensions de la
règle~\ref{eq:patt1} sont introduites~:

\begin{enumerate}

  \item Les expressions isolées du couple de plans alternants, de la
    forme $(l_1l_2 | l_2l_1)$, sont également prises en considération.

  \item Par ailleurs, dans les cas où deux motifs $(l_1, l_2)$ et
    $(l_1, l_3)$ partagent un même plan $l_1$, les plans $l_2$ et
    $l_3$ sont considérés comme équivalents~: en effet, de telles
    situations se produisent lors des scènes dialoguées quand un
    personnage est filmé de deux points de vue différents alors que
    son interlocuteur n'est filmé que d'un seul point de vue. La
    Figure~\ref{fig:motif_2} illustre une telle situation.

\end{enumerate}

Le pourcentage de parole couverte par les motifs de plans alternants
est donné dans le tableau~\ref{table:motif_stat}. Est également
reporté le temps de parole moyen couvert par une séquence de plans
alternants, ainsi que le nombre moyen de locuteurs impliqués dans une
séquence. Ces données sont indiquées à la fois pour les séquences
capturées par l'expression régulière $r$ et pour les séquences
capturées en utilisant sa version enrichie.

\begin{table}[h]
  \caption{\label{table:motif_stat}{\it Motifs de plans et parole: données statistiques}}
  \vspace{2mm}
  \centering
  \begin{tabular}{|c||c|c|c|}
    \hline
    & \textbf{couverture (\%)} & \textbf{temps parole/motif (s.)} & \textbf{nb.
      locuteurs/motif} \\
    \hline
    $r$ & 49.51 & 11.07 & 1.77 \\
    \hline
    ext. $r$ & \textbf{51.99} & \textbf{20.90} & \textbf{1.86} \\
    \hline
  \end{tabular}
\end{table}

Comme on peut le noter, les séquences de plans alternants couvrent un
peu plus de la moitié (51.99\%) du temps de parole total contenu dans
les épisodes de notre corpus.

D'autre part, 69.85\% des séquences contiennent deux locuteurs, 8.09\%
trois et 22.06\% seulement un~: les cas de séquences impliquant
uniquement un locuteur se produisent essentiellement pour les courtes
scènes dialoguées, où il peut arriver qu'un des deux personnages reste
silencieux.

Enfin, 97.96\% des personnages parlant au moins 5\% du temps total
sont impliqués dans de telles séquences.

\section{Détection locale des locuteurs}

\label{sec:seg_locale}

La segmentation du flux en locuteurs est accomplie en deux temps~:
d'abord localement en regroupant en classes de locuteurs l'ensemble
des segments parlés inclus dans une séquence de plans alternants~;
dans un deuxième temps, les locuteurs conjecturés dans chaque scène
dialoguée sont comparés afin de regrouper les locuteurs récurrents
d'une scène à l'autre.

\subsection{Descripteurs acoustiques}

Facilement disponibles, les plages des sous-titres originaux sont ici utilisées
pour estimer les limites temporelles des segments parlés. Malgré un
léger décalage temporel, d'amplitude variable, le sous-titre originale
transcrit fidèlement le segment parlé et coïncide avec ses limites
temporelles. Lorsque le temps de latence apparaissait trop important,
les limites temporelles des segments parlés concernés ont été
manuellement ajustées.

Par ailleurs, une même plage de sous-titre transcrit les propos d'un
seul locuteur. Dans les rares cas où les propos de deux locuteurs
distincts sont regroupés sur un seul sous-titre, les limites des tours
de parole sont explicitement indiquées et le sous-titre peut-être
subdivisé.

Chacun des segments parlés délimités d'après les plages de sous-titres
ne peut donc être affecté qu'à un seul locuteur. Ainsi, le pré-requis
de la plupart des systèmes de segmentation en locuteurs, à savoir la
détection des points de changement entre locuteurs dans le flux de
parole, peut être contourné pour mieux se concentrer sur la phase de
regroupement des segments en classes de locuteurs.

Pour caractériser les segments parlés, on extrait du signal audio les
19 premiers coefficients cepstraux, l'énergie ainsi que leurs dérivées
première et seconde, pour un total de 60 composantes.

Afin d'extraire l'information acoustique pertinente pour caractériser
le locuteur correspondant, on associe alors à chaque segment parlé un
$i$-vecteur \cite{dehak2011front} de 60 composantes, après
entraînement d'un modèle du monde de 512 composantes et apprentissage
d'une matrice de variabilité totale sur un sous-ensemble du corpus.

\subsection{Regroupement hiérarchique ascendant}

\label{ssec:part_local}

Un première étape de regroupement hiérarchique ascendant (regroupement
agglomératif) est accomplie localement, dans les limites de chacune
des scènes dialoguées définies selon la méthode décrite dans la
sous-section~\ref{ssec:motif}.

L'ensemble des segments parlés $\mathbf{u}(l_1, l_2)$ couverts par le
motif de plans alternants $(l_1, l_2)$ est alors partitionné en
classes de locuteurs selon les modalité suivantes~:

\begin{itemize}

  \item La distance de Mahalanobis est choisie comme mesure de
    similarité entre les $i$-vecteurs associés aux segments parlés.

    La matrice de covariance utilisée pour évaluer la distance de
    Mahalanobis entre $i$-vecteurs est la matrice de covariance
    intra-classe mentionnée dans~\cite{bousquet2011intersession},
    apprise sur un sous-ensemble du corpus et dont la formule est
    donnée par~:

    \begin{equation}
      W = \frac{1}{n} \sum_{s=1}^S \sum_{i=1} ^{n_s} (\mathbf{u_i^s} -
      \overline{\mathbf{u_s}}) (\mathbf{u_i^s} -
      \overline{\mathbf{u_s}})^T
    \end{equation}

    où $n$ désigne le nombre de segments parlés du corpus
    d'apprentissage, $S$ le nombre de locuteurs et $n_s$ le nombre de
    segments proférés par le locuteur $s$~; $\overline{\mathbf{u_s}}$
    est la moyenne des $i$-vecteurs associés aux segments parlés du
    locuteur $s$ et $\mathbf{u_i^s}$ désigne le $i$-vecteur associé à
    au $i$-ème segment parlé du locuteur $s$.

  \item Lors du processus agglomératif, le critère d'agrégation de
    Ward est utilisé pour évaluer la distance $\Delta I(c, c')$ entre
    les classes $c$ et $c'$, selon la formule suivante~:

    \begin{equation}
      \Delta I(c, c') = \frac{m_c m_{c'}}{m_c + m_{c'}} d^2 (g_c,
      g_{c'})
      \label{eq:ward}
    \end{equation}

    où $m_c$ et $m_{c'}$ désignent les masses des deux classes $c$  et
    $c'$, $g_c$ et $g_{c'}$ leurs centres de gravité respectifs et
    $d(g_c, g_{c'})$ la distance entre les deux centres de gravité.

  \item Enfin, la méthode Silhouette est utilisée pour couper le
    dendogramme issu de l'algorithme agglomératif à un niveau optimal
    et obtenir la partition finale des segments en classes de
    locuteurs. Décrite dans~\cite{rousseeuw1987silhouettes}, la
    méthode Silhouette permet d'associer à chaque partition possible
    du jeu de données un score normalisé entre -1 et 1~: pour une
    partition donnée, si certaines instances apparaissent plus proches
    du centre d'un autre classe que du centre de leur propre classe, le
    score tend à décroître, et à croître si elles sont plus proches du
    centre de leur propre classe que du centre d'une autre classe.

\end{itemize}

\section{Regroupement global sous contrainte}

\label{sec:seg_globale}

Un fois les segments parlés regroupés en classes de locuteurs dans
chaque scène dialoguée, une seconde phase de regroupement a lieu pour
regrouper au sein d'une même classe les locuteurs récurrents d'une
scène à l'autre.

L'ensemble des segments parlés assignés lors de la première étape de
regroupement à un même locuteur sont alors concaténés en un seul
segment audio, qu'on associe après paramétrisation acoustique à un
unique $i$-vecteur de dimension 60, caractéristique du locuteur
conjecturé localement.

L'ensemble de $i$-vecteurs qui en résulte est alors partitionné après
application d'un algorithme agglomératif selon les mêmes modalités que
celles décrites dans la sous-section~\ref{ssec:part_local}~: distance
de Mahalanobis fondée sur une matrice de covariance intra-classe,
critère d'agrégation de Ward, méthode de partionnement optimal
Silhouette.

Toutefois, l'information de structure acquise après la première phase
de partionnement est propagée à chaque étape de cette seconde étape de
regroupement~: dans la recherche des locuteurs récurrents, on doit en
effet empêcher que les locuteurs impliqués dans un même dialogue ne
soient regroupés dans la même classe de locuteurs.

On peut donc contraindre cette seconde phase de regroupement en
propageant à chaque étape du processus agglomératif ce
que~\cite{davidson2009using} décrit comme une contrainte de type
\og{}impossible-à-relier\fg{} (\textit{cannot-link}) qui interdit aux
locuteurs simultanément impliqués dans un même dialogue d'être
regroupés.

L'intégration d'une telle contrainte au processus de regroupement est
réalisée selon les modalités suivantes~:

\begin{itemize}

\item Dans la matrice de distance entre les $i$-vecteurs associés aux
  classes locales de locuteurs, la distance entre deux vecteurs et
  définie comme infinie si les deux locuteurs correspondants
  apparaissent conjointement dans le même dialogue~:

  \begin{equation}
    d(\mathbf{s}, \mathbf{s'}) = +\infty \Leftrightarrow \exists
    (l_1, l_2), \ \mathbf{u}(\mathbf{s}) \cup
    \mathbf{u}(\mathbf{s'}) \subseteq \mathbf{u}(l_1, l_2)
    \label{eq:rl1}
  \end{equation}

  où $(l_1, l_2)$ désigne un motif dialogué, $\mathbf{u}(l_1, l_2)$,
  l'ensemble de segments parlés couverts par le motif $(l_1, l_2)$ et
  $\mathbf{u}(\mathbf{s})$ l'ensemble des segments attribués au
  locuteur $\mathbf{s}$ après la phase de regroupement local .

\item Lors de la ré-évaluation des distances entre classes après
  chaque itération du processus agglomératif, la contrainte
  \textit{cannot-link} est propagée en situant à une distance infinie
  deux classes $c$ et $c'$ qui contiendraient respectivement des
  $i$-vecteurs incompatibles.

  \begin{equation}
    \Delta I(c, c') = +\infty \Leftrightarrow \exists (\mathbf{s},
    \mathbf{s'}) \in c \times c', \ d(\mathbf{s}, \mathbf{s'}) =
    +\infty
    \label{eq:rl2}
  \end{equation}

  où $\mathbf{s}$ et $\mathbf{s'}$ désignent les $i$-vecteurs
  correspondant à des classes de locuteurs conjecturées localement.

\end{itemize}

L'observation des règles~\ref{eq:rl1} et~\ref{eq:rl2} empêche deux
locuteurs impliqués dans le même dialogue d'être regroupés lorsque les
deux classes les plus proches sont fusionnées à chaque étape du
processus agglomératif.

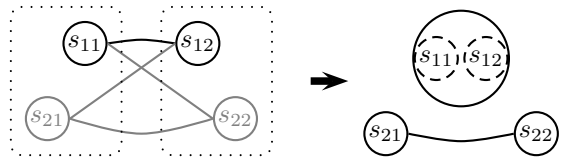
\begin{figure}[htb]
  \begin{minipage}[b]{1.0\linewidth}
    \centering
    \begin{pspicture}(8, 2)
      \psline[linewidth=.1]{->}(4, 1)(4.5, 1)
      \pscircle[linecolor=gray](0.5, 0.5){.3}
      \rput(0.5, 0.5){\textcolor{gray}{$s_{21}$}}
      \pscircle(1, 1.5){.3}
      \rput(1, 1.5){$s_{11}$}
      \pscircle[linecolor=gray](3, 0.5){.3}
      \rput(3, 0.5){\textcolor{gray}{$s_{22}$}}
      \pscircle(2.5, 1.5){.3}
      \rput(2.5, 1.5){$s_{12}$}
      \pscurve[linecolor=gray]{-}(0.8, 0.5)(1.75, 0.3)(2.7, 0.5)
      \pscurve{-}(1.3, 1.5)(1.75, 1.55)(2.2, 1.5)
      \psline[linecolor=gray]{-}(0.8, 0.5)(2.2, 1.5)
      \psline[linecolor=gray]{-}(1.3, 1.5)(2.7, 0.5)
      \psframe[framearc=.2, linestyle = dotted](0, 0)(1.5, 2)
      \psframe[framearc=.2, linestyle = dotted](2, 0)(3.5, 2)
      \pscircle(6, 1.3){.65}
      \pscircle[linestyle=dashed](5.67, 1.3){.3}
      \rput(5.67, 1.3){$s_{11}$}
      \pscircle[linestyle=dashed](6.33, 1.3){.3}
      \rput(6.33, 1.3){$s_{12}$}
      \pscircle(5, 0.3){.3}
      \rput(5, 0.3){$s_{21}$}
      \pscircle(7, 0.3){.3}
      \rput(7, 0.3){$s_{22}$}
      \pscurve{-}(5.3, 0.3)(6, 0.2)(6.7, 0.3)
    \end{pspicture}
  \end{minipage}
  \caption{{\it Première itération de l'algorithme agglomératif contraint}}
  \vspace{0.5cm}
  \label{clust}
\end{figure}

La Figure~\ref{clust} illustre la première itération de l'algorithme
agglomératif contraint ainsi que la manière dont la
contrainte~\textit{cannot-link} se propage.

On se place dans un cadre simplifié où seules deux scènes dialoguées
sont prises en considération~: chacune d'entre elles est délimitée par
des traits pointillés. On suppose d'autre part avoir conjecturé deux
locuteurs dans chacune d'entre elles~: $s_{ij}$ représente le
$i$-ème locuteur de la $j$-ème scène. L'éventuelle arête reliant deux
locuteurs figure la distance entre les deux $i$-vecteurs associés et
l'absence de liens correspond à une distance infinie.

L'algorithme va commencer par regrouper les deux locuteurs les plus
proches, soit $s_{11}$ et $s_{12}$, le groupe résultant pouvant
correspondre à un même locuteur récurrent d'une scène à
l'autre. L'absence de liens entre le locuteur récurrent et ses
interlocuteurs respectifs dans les deux scènes montre alors comme la
contrainte~\textit{cannot-link} est héritée par la classe issue de
chaque itération de l'algorithme agglomératif~: le seul regroupement
possible à la seconde itération de l'algorithme (non représentée sur
la Figure~\ref{clust}) serait de fusionner $s_{21}$ et $s_{22}$. La
propagation de cette contrainte vise à prévenir les regroupements
précoces d'interlocuteurs dans une même classe~: la musique de fond
peut en effet, par exemple, recouvrir la variabilité entre locuteurs
et provoquer prématurément un tel regroupement si aucune contrainte
n'est posée.

D'autre part, on voit comment le processus agglomératif est bloqué par
la propagation de ce type de contrainte~: dans le petit exemple
représenté sur la Figure~\ref{clust}, seules deux itérations de
l'algorithme sont possibles, avec au minimum deux classes de locuteurs
au terme du processus au lieu d'une seule en l'absence de contraintes.

La Figure~\ref{clust} met en regard les dendogrammes issus de deux
algorithmes agglomératifs, non-contraint et contraint, appliqués aux
mêmes données. La partie haute de la figure représente le dendogramme
issu du regroupement ascendant des classes de locuteurs~: le processus
peut être poursuivi jusqu'à ce qu'il ne reste plus qu'une classe
unique rassemblant toutes les instances~; en revanche, dans le cas du
regroupement effectué sous la contrainte de dissociation des
interlocuteurs, le processus agglomératif est prématurément bloqué et
débouche sur cinq dendogrammes disjoints, correspondant à cinq classes
incompatibles~: chaque classe contient au moins un locuteur en
dialogue direct avec au moins un locuteur de chacune des quatre autres
classes. On est donc en présence d'au moins cinq locuteurs récurrents.

\begin{figure}[htb]
  \begin{minipage}[b]{1.0\linewidth}
    \centering
    \centerline{\includegraphics[width=8cm]{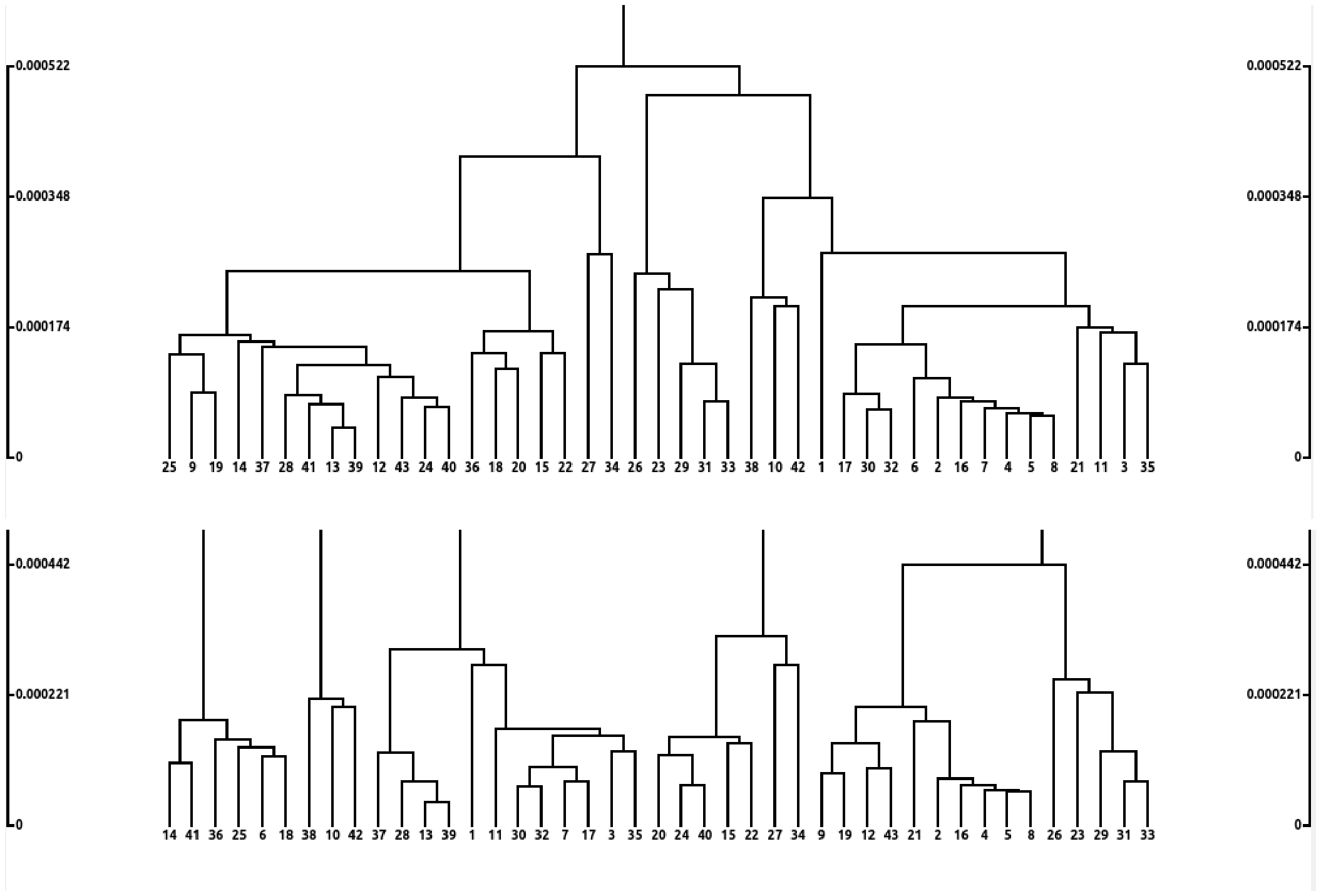}}
  \end{minipage}
  \caption{{\it Dendogrammes obtenus par regroupement agglomératif de
      locuteurs locaux, non-contraint (haut); contraint (bas)}}
  \label{dendo}
\end{figure}

Le partitionnement final en classes de locuteurs est alors obtenu en
partitionnant chacun des dendogrammes selon la méthode Silhouette
décrite dans la sous-section~\ref{ssec:part_local}.

Cette propriété d'irréductibilité~\cite{davidson2009using} d'une
partition, où les classes obtenues ne peuvent plus être réunies sans
violer une contrainte de dissociation des instances, permet ainsi de
contourner en partie le problème critique du partitionnement optimal
du dendogramme en bloquant prématurément le processus d'agglomération.

Toutefois, cette seconde phase de regroupement reste dépendante de la
première phase de partitionnement des segments parlés en classes
locales de locuteurs. Si les propos d'un même locuteur sont à tort
répartis sur deux classes de locuteurs disjointes pendant la première
phase, ils ne pourront plus être regroupés lors de la deuxième étape,
les deux classes étant désormais considérées comme
incompatibles. Cependant, même lors d'une regroupement non contraint,
ces segments, dissociés au niveau local, ne serait regroupés que
tardivement au niveau global, éventuellement après que le
partitionnement soit jugé optimal.

D'autre part, une erreur peut être commise lors de la phase de
regroupement global~: pour des locuteurs $a$, $b$, et $c$, et en
posant, en reprenant l'exemple de la Figure~\ref{clust}, $s_{11} := a$,
$s_{21} := c$, $s_{12} := b$, $s_{22} := a$, la première itération de
l'algorithme sous contrainte, en regroupant $a$ et $b$, aura pour
effet de placer le locuteur $a$ dans deux dendogrammes disjoints,
empêchant toute fusion ultérieure des deux occurrences de $a$, et
dégradera la couverture des classes de locuteurs. Toutefois, dans le
cas d'un partitionnement non contraint, les deux occurrences du
locuteur $a$, non regroupées après la première itération, ne le
seraient probablement que tardivement, éventuellement après que le
partitionnement optimal aura été atteint.

Enfin, pour un ensemble donné d'instances entre lesquelles sont posées
des contraintes de séparation à respecter lors du regroupement, le
nombre final de classes de la partition irréductible semble dépendant
de l'ordre dans lequel les regroupements deux à deux ont
lieu~\cite{davidson2009using}. Toutefois, comme on le verra dans la
section~\ref{sec:expe} qui suit, le nombre final de classes obtenues
en partitionnant les dendogrammes issus du regroupement contraint
est apparu expérimentalement offrir une approximation raisonnable du
nombre réel de locuteurs impliqués.

\section{Expériences et résultats}

\label{sec:expe}

\subsection{Corpus}

Notre corpus est constitué d'épisodes des premières saisons de trois
séries~: \textit{Breaking Bad} (abrégé \textit{bb}), \textit{Game of
  Thrones} (\textit{got}) et \textit{House of Cards}
(\textit{hoc}). Nous avons manuellement annoté trois épisodes de
chaque série en indiquant les coupes de plans, les plans similaire,
les segments parlés ainsi que les locuteurs associés.

Le temps de parole total contenu dans ces neuf épisodes représente un
peu plus de trois heures (3:12).

Un sous-ensemble de six épisode (désigné \textsc{dev}) a été utilisé à
des fins de développement (seuils de coupes et de similarité entre
plans) ou d'apprentissage (modèle \textsc{gmm/ubm}, matrice de
variabilité totale, matrice de covariance intra-classe). Les trois
épisodes restants (\textsc{test}) ont été utilisés à des fins de test.

\subsection{Détection des coupes et des similitudes entre plans}

L'évaluation de la méthode de détection des coupes repose sur une
F1-mesure \cite{boreczky1996comparison} fondée sur le rappel (taux de
coupes retrouvées parmi les coupes pertinentes) et la précision (taux
de coupes pertinentes parmi celles qui ont été conjecturées). Pour la
tâche de similarité entre plans, un F1-score analogue est utilisé~:
pour chaque plan, on compare la liste des plans conjecturés comme lui
étant similaires à celle des plans annotés comme semblables~: si
l'intersection des deux listes est non vide, le plan est considéré
comme correctement apparié. Les résultats sur les ensembles de
\textsc{dev} et de \textsc{test} sont présentés dans le
Tableau~\ref{table:res_plans}.

\begin{table}[h]
  \caption{\label{table:res_plans}{\it Résultats obtenus pour la détection de
      coupes et de similitudes entre plans}}
  \vspace{2mm}
  \centering
  \begin{tabular}{|c|c||ccc|}
    \hline
    & \textbf{coupes} & \multicolumn{3}{|c|}{\textbf{similitudes}} \\
    \cline{2-5}
    & F1-score & précision & rappel & F1-score \\
    \hline
    \hline
    \textit{bb-1} & 0.93 & 0.88 & 0.81 & 0.84 \\
    \hline
    \textit{bb-2} & 0.99 & 0.90 & 0.83 & 0.86 \\
    \hline
    \textit{got-1} & 0.97 & 0.88 & 0.84 & 0.86 \\
    \hline
    \textit{got-2} & 0.98 & 0.89 & 0.90 & 0.90 \\
    \hline
    \textit{hoc-1} & 0.99 & 0.91 & 0.92 & 0.92 \\
    \hline
    \textit{hoc-2} & 0.98 & 0.93 & 0.97 & 0.95 \\
    \hline
    \hline
    moy. \textsc{dev} & \textbf{0.97} & 0.90 & 0.88 & \textbf{0.89} \\
    \hline
    \hline
    \textit{bb-3} & 0.98 & 0.83 & 0.84 & 0.83 \\
    \hline
    \textit{got-3} & 0.99 & 0.92 & 0.89 & 0.91 \\
    \hline
    \textit{hoc-3} & 0.99 & 0.98 & 0.96 & 0.97 \\
    \hline
    \hline
    moy. \textsc{test} & \textbf{0.99} & 0.91 & 0.90 & \textbf{0.90} \\
    \hline
  \end{tabular}
\end{table}

Le résultats obtenus pour les deux tâches relevant du traitement de
l'image apparaissent élevés~: le F1-score obtenu lors de la détection
des images similaires en particulier (0.90) permet de déterminer avec
confiance les limites des passages dialogués au sein desquels la
segmentation en locuteurs sera effectuée.

\subsection{Segmentation locale en locuteurs}

Le taux d'erreur utilisé pour évaluer la segmentation locale en
locuteurs est déterminé indépendamment pour chaque scène dialoguée,
avant d'être estimé globalement par la somme des taux d'erreurs locaux
pondérés par la durée de chaque scène (\textit{single-show Diarization
  Error rate}, tel que défini dans~\cite{rouvier2013open}). Les
résultats sont reportés dans le Tableau~\ref{table:res_loc}, à la fois
en utilisant les similitudes de plans de référence (\textit{entrée
  réf.}) et les similitudes de plans déterminées automatiquement
(\textit{entrée auto.}). D'autre part, à des fins de comparaison, le
regroupement agglomératif (désigné \textit{ra}) est comparée à un
principe de regroupement \og{}naïf\fg{} fondé sur la seule alternance
des plans~: chaque segment parlé est étiqueté selon le plan qui
s'affiche, en supposant que l'alternance des plans coïncide avec
l'alternance des locuteurs.

\begin{table}[h]
  \caption{\label{table:res_loc}{\it Segmentation locale en locuteurs~:
      \textsc{der}}}
  \vspace{2mm}
  \centering
  \begin{tabular}{|c|cc||cc|}
    \hline
    & \multicolumn{2}{|c||}{\textbf{entrée auto.}} &
    \multicolumn{2}{|c|}{\textbf{entrée réf.}} \\
    \cline{2-5}
    & \textit{naïf} & \textsc{ra} & \textit{naïf} & \textsc{ra} \\
    \hline
    \hline
    \textit{bb-1} & 30.26 & \textbf{19.11} & 22.81 & 21.00 \\
    \hline
    \textit{bb-2} & 22.06 & 22.51 & 19.78 & \textbf{19.14} \\
    \hline
    \textit{got-1} & 22.16 & 23.70 & 19.46 & \textbf{15.78} \\
    \hline
    \textit{got-2} & 26.19 & 18.78 & 22.80 & \textbf{16.61} \\
    \hline
    \textit{hoc-1} & 17.23 & 13.36 & 16.31 & \textbf{11.84} \\
    \hline
    \textit{hoc-2} & 30.66 & \textbf{18.18} & 31.87 & 19.12 \\
    \hline
    \hline
    moy. \textsc{dev} & 24.76 & 19.27 & 22.17 & \textbf{17.25} \\
    \hline
    \hline
    \textit{bb-3} & 40.45 & 21.15 & 24.31 & \textbf{12.15} \\
    \hline
    \textit{got-3} & 33.45 & 17.43 & 35.43 & \textbf{12.80} \\
    \hline
    \textit{hoc-3} & 24.44 & 12.83 & 22.95 & \textbf{12.82} \\
    \hline
    \hline
    moy. \textsc{test} & 32.78 & 17.14 & 27.56 & \textbf{12.59} \\
    \hline
  \end{tabular}
\end{table}

Les résultats obtenus par le regroupement des segments parlés sur un
critère purement acoustique apparaissent supérieurs à ceux obtenus par
la méthode naïve fondée sur la seule alternance des plans.

Par ailleurs, l'automatisation de l'étape précédente de détection des
plans similaires n'affecte qu'en partie les résultats de la
segmentation des locuteurs (augmentation du taux d'erreur de près de
5\% sur les trois épisodes du \textsc{test}, mais seulement de 2\% sur les six
épisodes du \textsc{dev}).

\subsection{Segmentation globale sous contrainte}

La tableau~\ref{table:res_glob} regroupe les résultats obtenus lors de
la seconde étape de regroupement global des locuteurs récurrents. Les
résultats sont donnés à la fois en prenant en entrées les locuteurs
conjecturés lors de la phase de segmentation locale (\textit{entrée
  auto.}) et les locuteurs de référence (\textit{entrée réf.}). Dans
les deux cas, les résultats sont donnés pour la version non contrainte
du regroupement (désignée \textit{2S}) et pour sa version contrainte
(\textit{cnt. 2S}). Les résultats obtenus sont comparés à ceux obtenus
par deux outils standards de segmentation en locuteurs fondés sur des
algorithmes hiérarchiques (dénotés \textsc{lia} et \textsc{lium})
auxquels on fournit en entrée les segments parlés de référence
couverts par les dialogues capturés selon les modalités décrites dans
la section~\ref{sec:dialogues}.

\begin{table}[h]
  \caption{\label{table:res_glob}{\it \textsc{der} Segmentation globale en
      locuteurs: \textsc{der}}}
  \vspace{2mm}
  \centering
  \begin{tabular}{|c|cc||cc||cc|}
    \hline
    & \multicolumn{2}{|c||}{\textbf{entrée auto.}} &
    \multicolumn{2}{|c||}{\textbf{entrée réf.}} &
    \multicolumn{2}{|c|}{\textbf{segments réf.}} \\
    \cline{2-7}
    & \textit{2S} & \textit{cnt. 2S} & \textit{2S} & \textit{cnt. 2S}
    & \textsc{lia} & \textsc{lium} \\
    \hline
    \hline
    \textit{bb-1} & 51.36 & 56.00 & 52.66 & \textbf{48.10} & 72.06 & 67.21 \\
    \hline
    \textit{bb-2} & \textbf{41.83} & 65.07 & 58.76 & 49.49 & 77.03 & 76.79 \\
    \hline
    \textit{got-1} & 70.13 & \textbf{52.79} & 70.67 & 53.87 & 65.57 & 58.49 \\
    \hline
    \textit{got-2} & 67.28 & \textbf{38.85} & 70.32 & 41.24 & 65.29 & 60.80 \\
    \hline
    \textit{hoc-1} & \textbf{50.04} & 55.61 & 52.70 & 52.15 & 60.26 & 62.37 \\
    \hline
    \textit{hoc-2} & 64.91 & 56.40 & 63.65 & \textbf{37.09} & 67.05 & 59.00 \\
    \hline
    \hline
    moy. & 57.59 & 54.11 & 61.46 & \textbf{46.99} & 67.88 & 64.11 \\
    \hline
    \hline
    \textit{bb-3} & 60.41 & \textbf{33.94} & 59.22 & 42.64 & 60.61 & 55.56 \\
    \hline
    \textit{got-3} & 74.71 & \textbf{49.31} & 70.34 & 63.17 & 61.33 & 52.89 \\
    \hline
    \textit{hoc-3} & \textbf{57.68} & 59.87 & 67.52 & 67.41 & 70.55 & 67.05 \\
    \hline
    \hline
    moy. & 64.13 & \textbf{47.71} & 65.69 & 57.74 & 64.16 & 58.50 \\
    \hline
  \end{tabular}
\end{table}

Bien qu'encore élevé, le taux d'erreur est en général réduit en
propageant dans le processus de regroupement la contrainte de
dissociation des interlocuteurs. En bloquant le processus de
regroupement au moment où il devient irréductible, l'information de
structure ainsi propagée permet de contourner en partie le problème
délicat du niveau optimal où couper le dendogramme issu du processus
agglomératif.

Le Tableau~\ref{table:nspk} donne, pour chacune des méthodes de
segmentation en locuteurs présentée dans le
Tableau~\ref{table:res_glob}, le nombre moyen de locuteurs par
épisode, en regard du nombre réel, pour chacune des trois séries de
notre corpus.

\begin{table}[h]
  \caption{\label{table:nspk}{\it Nombre moyen de locuteurs par
      épisode}}
  \vspace{2mm}
  \centering
  \begin{tabular}{|c|c||cc||cc|}
    \hline
    & \textbf{réf.} & \textit{2S} & \textit{cnt. 2S} & \textsc{lia} &
    \textsc{lium} \\
    \hline
    \hline
    \textit{bb} & \textbf{10.3} & 7.3 & \textbf{11} & 6 & 25.7 \\
    \hline
    \textit{got} & \textbf{25.3} & 4.7 & 15.7 & 9.3 & \textbf{24} \\
    \hline
    \textit{hoc} & \textbf{20.7} & 3.7 & \textbf{24} & 6 & 27 \\
    \hline
  \end{tabular}
\end{table}

Comme on peut le constater, deux des systèmes mentionnés, notre
système non contraint (\textit{2S}) et \textsc{lia}, ont tendance à
sous-estimer le nombre total de locuteurs par épisode en coupant le
dendogramme à un niveau élevé. Inversement, le système désigné par
\textsc{lium} a tendance à surestimer le nombre de locuteurs impliqués
en opérant une coupe basse de l'arbre de regroupement. L'approche
fondée sur la contrainte de dissociation des interlocuteurs
(\textit{cnt. 2S}) débouche en revanche sur une partition irréductible
qui permet d'estimer raisonnablement le nombre de locuteurs.

\section{Conclusion}

Dans ce papier, nous avons proposé de procéder à la segmentation de
séries télévisées en locuteurs en nous appuyant sur la structure
narrative qui les sous-tend. En détectant les plans similaires, des
séquences d'alternance de plans typiques des scènes dialoguées peuvent
être délimitées et une première phase de regroupement des segments
parlés en classes de locuteurs peut avoir lieu dans les limites de
chacune des scènes isolées. Une seconde étape de regroupement, visant
à détecter les locuteurs récurrents, est alors appliquée aux classes
de locuteurs localement conjecturées~: à chaque itération du processus
d'agglomération, une contrainte de séparation des interlocuteurs
impliqués dans une même scène est propagée. Le processus de
regroupement se trouve alors bloqué avant que toutes les instances ne
puissent être groupées. La partition irréductible qui en résulte
permet d'estimer plus facilement le nombre total de locuteurs
impliqués.

Il resterait à mener une étude plus systématique sur la manière dont
l'ordre dans lequel les instances sont groupées sous une contrainte de
séparation affecte le nombre final de sous-ensembles de la partition
irréductible.

D'autre part, c'est essentiellement lors de la seconde phase de
regroupement que le taux d'erreur se dégrade sensiblement~: même en
prenant en entrée les locuteurs de référence impliqués dans les
différentes scènes dialogués, certains des regroupements effectués
apparaissent lourdement erronés. La variabilité dans l'environnement
acoustique d'une scène à l'autre semble induire nombre de ces
erreurs. Il faudrait donc étudier plus systématiquement, et
éventuellement isoler, les sources de variabilités que peuvent
constituer la musique de fond ou les bruitages.

\acknowledgements{Ce travail a été en partie soutenu par le projet
  \textsc{contnomina} (\textsc{anr}-07-240) de l'Agence Nationale de
  la Recherche, ainsi que par la Fédération de Recherche
  \textit{Agorantic}, Université d'Avignon et des Pays de Vaucluse.}

\bibliography{base}

\end{document}